\documentclass[
showkeys,	
reprint,	
amsmath,	
amssymb,	
amsfonts,	
aps,		
pre,%
groupedaddress,
superscriptaddress,
a4paper,%
nofootinbib,
eprint,		
final,		
preprintnumbers, 
]{revtex4-2}

\usepackage[tbtags]{mathtools}
\usepackage{amsmath,amsfonts,mathdots,amssymb,yfonts,calc}
\usepackage{graphicx,graphics,xcolor}
\usepackage{tabularx}
\usepackage{subcaption}

\usepackage{booktabs}
\usepackage{siunitx}
\usepackage{float}
\usepackage{tabularray}
\usepackage{multirow}
\usepackage{physics}
\usepackage{dcolumn}
\usepackage[utf8]{inputenc}  
\usepackage[T1]{fontenc} 
\usepackage{natbib}
\usepackage{booktabs}
\usepackage{hyperref}
\hypersetup{
    colorlinks=true,
    linkcolor=blue,
    filecolor=magenta,      
    urlcolor=blue!80,
    citecolor=cyan,
    linktocpage=true,
    breaklinks=false,
}
\begin{document}
\title{Complexity Condensation Through Adaptive Information Exchange}
\author{Vinesh Vijayan}
\thanks{Corresponding author: vinesh.physics@rathinam.in}
\affiliation{Department of Science and Humanities, Rathinam Technical Campus, Coimbatore, India -641021}
\author{R Prakash}
\affiliation{Department of Science and Humanities, Rathinam Technical Campus, Coimbatore, India -641021}
\author{D Kaleeswaran}
\affiliation{Department of Computer Science Engineering, Rathinam Technical Campus, Coimbatore, India -641021}

\date{\today}

\begin{abstract}
An emergent complexity field governing information exchange is the central theme of this work. To explore this idea, we propose a model of an adaptive dynamical network in which both the interaction weights and the adaptive coupling strengths are determined by finite-time information production rates that quantify the dynamical complexity of individual subsystems. Collective organization in complexity space emerges through a feedback mechanism between the microscopic dynamics and the resulting complexity-dependent interactions. Using numerical simulations, we demonstrate the emergence of a phenomenon that we term \emph{complexity condensation}, in which subsystem complexities become strongly localized despite the absence of complete state synchronization. The degree of condensation is found to be maximal at an intermediate adaptation strength, reflecting a balance between selective information exchange and network fragmentation. These results reveal a mechanism for complexity-mediated self-organization in nonlinear systems driven by adaptive information exchange.

\end{abstract}

\keywords{}
\maketitle
\pagestyle{plain}

\section{\label{s1}Introduction}
A central challenge in nonequilibrium statistical physics and complex systems science is to understand how collective organization emerges from the interactions among large numbers of nonlinear units. The emergence of macroscopic order from microscopic dynamics through effective collective variables is a recurring theme across physical, biological, and social systems \cite{Anderson1972, Strogatz2001, Cross1993, Haken1983, Artime2022}. Representative examples include synchronization in coupled oscillator populations \cite{Winfree1967, Kuramoto1975, Strogatz2000, Acebron2005, Dorfler2013}, pattern formation in reaction–diffusion systems \cite{Turing1952,Cross1993,Murray1981,Kondo2010,Landge2020}, and self-organization in adaptive and evolving networks \cite{Bornholdt2000,Gross2006,GrossBlasius2008,
Vazquez2008,Bohme2011}. A common objective in these studies is the identification of coarse-grained descriptors that capture the essential dynamics of the system, since large-scale behavior cannot generally be inferred directly from microscopic state variables.

Lyapunov exponents occupy a central position among measures of nonlinear dynamics because they quantify local trajectory divergence and dynamical information production \cite{Oseledec1968,Benettin1980,Eckmann1985,Wolf1985,Aurell1997}. Positive Lyapunov exponents provide a direct signature of chaos and are closely related to entropy production through Pesin's theorem \cite{Pesin1977,Eckmann1985,Young2002}. Finite-time Lyapunov exponents (FTLEs) extend this framework by providing a local and temporally resolved characterization of dynamical activity, and have been widely employed in studies of chaotic transport, turbulence, spatiotemporal chaos, and complex networks \cite{Aurell1997,Boffetta2002,Haller2000,Shadden2005}. Despite their effectiveness in quantifying dynamical complexity, Lyapunov exponents are typically treated as diagnostic observables rather than active variables capable of influencing system evolution.

Studies of complex adaptive networks have demonstrated that interaction structures need not remain static but can coevolve with the dynamics taking place on them \cite{Bornholdt2000,Gross2006,GrossBlasius2008,Vazquez2008}. Such adaptive interactions generate a variety of emergent phenomena, including synchronization, self-organized criticality, clustering, and community formation \cite{Bornholdt2000,Gross2006,Ito2003,Zimmermann2004,Holme2006}. However, most adaptation rules are based on microscopic dynamical variables, phase differences, synchronization errors, or network topology. Comparatively little attention has been devoted to the possibility that interactions themselves may be regulated by measures of dynamical complexity. Such a feedback mechanism would provide a fundamentally different perspective in which complexity is not merely an outcome of the dynamics but also a factor shaping the evolution of the interaction network.

Recent information-theoretic approaches to complex systems have highlighted the roles of information production, information transfer, and collective computation in the emergence of large-scale organization \cite{Varley2024,Mediano2022,Rosas2018,Kohler2022,Bajic2024}. These developments motivate the hypothesis that subsystems with similar levels of dynamical complexity and information-processing capability may exchange information more efficiently than dissimilar ones. From this perspective, dynamical complexity may induce an effective interaction geometry that shapes information flow across the system. If such a mechanism operates, it could give rise to forms of collective organization fundamentally different from conventional synchronization phenomena.

Motivated by these considerations, we propose a framework in which dynamical complexity is treated as an active dynamical field rather than a passive observable. Each subsystem is assigned a finite-time complexity variable derived from its recent dynamical history. Information exchange occurs in an emergent complexity space, where an interaction kernel preferentially couples subsystems with similar complexity levels. The strength of information exchange adapts according to the mismatch between the complexity of a subsystem and that of its surrounding complexity environment. The resulting dynamics establishes a feedback loop in which microscopic dynamics generates complexity, complexity reshapes the interaction network, and the evolving interaction structure subsequently influences the microscopic dynamics.

The central question addressed in this work is whether adaptive information exchange can generate collective organization in complexity space. To quantify this phenomenon, we introduce the concept of \textit{complexity condensation}, defined as the concentration of subsystem complexities within a narrow region of complexity space. By analyzing the coupled evolution of microscopic dynamics, emergent complexity fields, and adaptive interactions, we demonstrate that complexity condensation emerges as a self-organized phenomenon and reaches its maximum at an intermediate adaptation strength. This behavior results from a nontrivial interplay between complexity-space locality and adaptive information filtering, leading to optimal condensation regimes and well-defined phase boundaries. Within this framework, emergent complexity fields simultaneously act as descriptors of dynamics, generators of complexity-space geometry, and regulators of information exchange. Our results identify a general mechanism through which large-scale organization can emerge from complexity-dependent interactions, while establishing connections between nonlinear dynamics, adaptive networks, and information-theoretic approaches to collective behavior.
\section{The Model}
We consider a network of $N$ nonlinear subsystems whose dynamics are governed by a chaotic map $f(x)$. The discrete-time evolution of the $i$th subsystem is given by
\begin{equation}
x_i(n+1)=\left[1-\epsilon_i(n)\right]f\left(x_i(n)\right)
+\epsilon_i(n)\sum_{j=1}^{N}W_{ij}(n)
f\left(x_j(n)\right),
\label{eq1}
\end{equation}
where the first term on the right-hand side represents the intrinsic dynamics of the subsystem, while the second term describes information exchange mediated through interactions in complexity space. The relative contributions of different information sources are determined by the interaction weights $W_{ij}(n)$, whereas the overall receptivity of the $i$th subsystem to external information is controlled by the adaptive coupling strength $\epsilon_i(n)$.

For the numerical investigations presented in this work, the local dynamics of each subsystem is chosen to be the logistic map\cite{May1976},
\begin{equation}
f(x)=rx(1-x),
\label{eq2}
\end{equation}
with the control parameter fixed at $r=4$, corresponding to fully developed chaotic dynamics. The choice of the logistic map is not essential to the theoretical framework. Rather, it serves as a simple and well-studied model of a nonlinear information-generating subsystem from which the emergent complexity field can be constructed.

A characteristic feature of complex systems is the separation between fast microscopic dynamics and the slower processes associated with collective organization. While the state variables of individual subsystems may fluctuate rapidly, the emergent large-scale behavior is often governed by coarse-grained quantities that encode the recent dynamical history of the system. Motivated by this observation, we introduce a complexity field, $\lambda_i(n)$, associated with each subsystem. This field is defined through a finite-time Lyapunov exponent,
\begin{equation}
\lambda_i(n)=
\frac{1}{T_{\lambda}}
\sum_{m=n-T_{\lambda}+1}^{n}
\ln\left|
f'\left(x_i(m)\right)
\right|,
\label{eq3}
\end{equation}
where $T_{\lambda}$ denotes the observation window. The quantity $\lambda_i(n)$ provides a local measure of dynamical instability and information production by characterizing the average rate of trajectory divergence over a finite time interval. Smaller values of $\lambda_i$ correspond to relatively regular dynamics, whereas larger values indicate stronger local unpredictability and enhanced information generation.

Since the complexity field is obtained through temporal averaging, it captures the persistent dynamical characteristics of each subsystem, whereas the microscopic state variable $x_i(n)$ evolves on much shorter time scales. At a given time step $n$, the network is characterized by the set

\begin{equation}
\lbrace
\lambda_1(n),\lambda_2(n),\ldots,\lambda_N(n)
\rbrace,
\label{eq4}
\end{equation}

which defines the instantaneous complexity landscape of the system. Within the present framework, this landscape acts as an emergent mesoscopic variable that not only characterizes the collective dynamical state of the network but also governs the geometry and strength of information exchange among subsystems. Consequently, the complexity field plays an active dynamical role rather than serving merely as a diagnostic observable.

The next step is to construct a complexity-dependent interaction kernel based on the emergent complexity field. In contrast to conventional approaches, where interactions are determined by similarity in microscopic states, the present framework assumes that interaction strengths are governed by similarity in dynamical complexity. This constitutes the central assumption underlying the construction of the interaction network. The underlying rationale is that subsystems possessing comparable levels of dynamical complexity are expected to process and exchange information more efficiently than dynamically dissimilar subsystems. Accordingly, the interaction weights are defined as
\begin{equation}
W_{ij}(n)=
\frac{
\exp\left[-\beta \left|\lambda_i(n)-\lambda_j(n)\right|\right]
}
{
\sum_{k=1}^{N}
\exp\left[-\beta \left|\lambda_i(n)-\lambda_k(n)\right|\right]
},
\label{eq5}
\end{equation}
where the normalization ensures that
\begin{equation}
\sum_{j=1}^{N}W_{ij}(n)=1.
\label{eq6}
\end{equation}
The parameter $\beta$ controls the locality of interactions in complexity space. For $\beta=0$, all subsystems interact with equal strength, and the network reduces to a mean-field coupling scheme. As $\beta$ increases, interactions become progressively localized in complexity space, causing information exchange to occur preferentially among subsystems with similar complexity levels. In the limit of large $\beta$, the interaction network becomes strongly concentrated around complexity-matched subsystems.

The outcome of these complexity-dependent interactions is the emergence of an effective geometry in complexity space. In this framework, subsystems interact according to their proximity in dynamical complexity rather than their separation in physical space. Using the interaction kernel defined above, we introduce the complexity environment of subsystem $i$ as
\begin{equation}
\bar{\lambda}_i(n)= \sum_{j=1}^{N} W_{ij}(n) \lambda_j(n),
\label{eq7}
\end{equation}
where $\bar{\lambda}_i(n)$ represents the weighted average complexity of the dynamically compatible neighborhood surrounding subsystem $i$. The weights $W_{ij}(n)$ ensure that the contribution of each subsystem is determined by its proximity to subsystem $i$ in complexity space. Consequently, $\bar{\lambda}_i(n)$ provides a local mesoscopic description of the complexity landscape experienced by subsystem $i$ and serves as the reference against which its complexity is compared in the adaptive interaction dynamics.

The interaction kernel determines which subsystems preferentially exchange information. However, the efficiency with which a subsystem incorporates external information may itself depend on its relationship to the surrounding complexity environment. We therefore assume that each subsystem continuously compares its own complexity level with the average complexity of its dynamically compatible neighborhood. Information exchange is expected to be most effective when a subsystem is dynamically compatible with its environment, whereas a significant complexity mismatch reduces its ability to incorporate external information. To capture this effect, the adaptive coupling strength is modeled as
\begin{equation}
\epsilon_i(n)=\epsilon_0\exp\left[-\alpha\left|\bar{\lambda}_i(n)-\lambda_i(n)\right|\right],
\label{eq8}
\end{equation}
where $\epsilon_0$ denotes the baseline coupling strength and the parameter $\alpha$ controls the strength of the adaptive filtering mechanism. When the complexity of a subsystem closely matches that of its surrounding complexity environment, the coupling approaches its maximum value $\epsilon_0$, allowing efficient information exchange. Conversely, increasing complexity mismatch suppresses the coupling exponentially, thereby reducing the influence of external information on the subsystem dynamics.

When $\alpha$ is small, and in particular for $\alpha=0$, all subsystems interact with the same baseline coupling strength $\epsilon_0$. As $\alpha$ increases, the coupling becomes progressively more sensitive to complexity mismatch. Consequently, the adaptive coupling introduces a competition between two opposing tendencies. On the one hand, moderate adaptive filtering suppresses complexity fluctuations by reducing interactions between dynamically incompatible subsystems. On the other hand, excessively strong filtering weakens information exchange throughout the network and can ultimately lead to dynamical fragmentation.

The complexity field therefore assumes a dual role within the present framework. It determines both the geometry of information exchange through the interaction weights $W_{ij}(n)$ and the efficiency of information transfer through the adaptive coupling strengths $\epsilon_i(n)$. As a result, the system is governed by the feedback loop

\begin{equation}
x_i(n)
;\rightarrow;
\lambda_i(n)
;\rightarrow;
\left[
W_{ij}(n),,\epsilon_i(n)
\right]
;\rightarrow;
x_i(n+1),
\label{eq9}
\end{equation}

in which microscopic dynamics generates the complexity field, the complexity field determines the interaction structure, and the resulting interactions subsequently influence the microscopic dynamics. This feedback mechanism constitutes the fundamental principle underlying the emergence of collective organization and complexity condensation in the present model.
\section{Complexity Condensation}

The central objective of the present work is to determine whether adaptive information exchange can generate collective organization in complexity space. To quantify this phenomenon, we examine the statistical distribution of the emergent complexity field across the network. At a given iteration step $n$, the state of the system is characterized by the set of complexity values defined in Eq.~(\ref{eq4}). These values give rise to a probability distribution $P(\lambda)$, representing the fraction of subsystems possessing a given complexity level.

In the absence of collective organization, $P(\lambda)$ is broadly distributed, indicating substantial heterogeneity in the dynamical instability and information production rates of individual subsystems. Adaptive interactions, however, can drive a significant fraction of subsystems to accumulate within a narrow region of complexity space, resulting in a sharply localized distribution centered around a characteristic complexity value. We refer to this concentration of complexity values as \textit{complexity condensation}. In this state, the complexity field exhibits a high degree of collective organization despite the persistence of microscopic dynamical fluctuations.

A comparison with conventional synchronization is instructive at this stage. Unlike synchronization, complexity condensation does not require the microscopic state variables of different subsystems to become identical. Subsystems may remain dynamically distinct while exhibiting similar levels of dynamical instability and information production. Consequently, the collective organization studied here occurs at the level of dynamical complexity rather than at the level of microscopic trajectories.

To quantify the degree of complexity condensation, we introduce the complexity variance,

\begin{equation}
M= \mathrm{Var}(\lambda)=\frac{1}{N}\sum_{i=1}^{N}\left[\lambda_i - \langle \lambda \rangle \right]^2,
\label{eq10}
\end{equation}

where

\begin{equation}
\langle \lambda \rangle=\frac{1}{N}\sum_{i=1}^{N}\lambda_i
\label{eq11}
\end{equation}

denotes the average complexity of the network. Small values of $M$ indicate that the subsystem complexities are concentrated within a narrow region of complexity space, corresponding to a highly condensed state. Conversely, large values of $M$ reflect broad complexity distributions and a high degree of heterogeneity across the network and therefore $M$ acts as an order parameter of the system.

As a matter of convenience, we also introduce the width of the complexity distribution,
\begin{equation}
\sigma_{\lambda}=\sqrt{M},
\label{eq12}
\end{equation}
which provides a direct measure of the spread of complexity values across the network. Complexity condensation corresponds to the emergence of collective states characterized by a narrow complexity distribution. In the idealized limit of complete condensation, the distribution approaches
\begin{equation}
P(\lambda) \rightarrow \delta(\lambda-\lambda_c),
\label{eq13}
\end{equation}
where $\lambda_c$ denotes the characteristic complexity of the condensed state. In finite systems, condensation manifests itself as a pronounced reduction in both the complexity variance $M$ and the distribution width $\sigma_{\lambda}$. Consequently, these quantities serve as natural order parameters for quantifying the degree of collective organization in complexity space.

We introduce the \emph{Complexity Condensation Efficiency} (CCE), which quantifies the extent to which the complexity measure becomes localized relative to a reference state. Larger values of this metric indicate a more effective adaptive self-organization process. The CCE is defined as
\begin{equation}
\eta=\frac{M(0)}{M(\alpha)},
\label{eq14}
\end{equation}
where $M(0)$ denotes the variance of the complexity measure in the reference state, i.e., in the absence of adaptive feedback ($\alpha=0$), and $M(\alpha)$ represents the corresponding variance for a given adaptive feedback strength $\alpha$. Thus, $\eta$ provides a quantitative measure of the reduction in the spatial (or statistical) spread of the complexity measure induced by adaptive feedback. Consequently, larger values of $\eta$ correspond to a greater degree of complexity condensation and, therefore, more efficient adaptive self-organization.

\section{Numerical Validation}
The dependence of the complexity variance $M$ on the control parameters $\alpha$ and $\beta$ provides direct insight into the role of adaptive information exchange in shaping collective organization. Of particular interest are the parameter regimes in which $M$ attains its minimum value, corresponding to maximal complexity condensation. These regimes identify the conditions under which adaptive information exchange most effectively reorganizes the complexity landscape of the network. Throughout this work, the emergence of collective organization in complexity space, as quantified by complexity condensation, serves as the primary observable for characterizing the system's macroscopic behavior.

In the following, numerical simulations are performed for a network of $N=200$ maps with $r=4$ and $\epsilon_0=0.4$. Statistical quantities are averaged over $100$ independent realizations, and asymptotic states are evaluated after discarding an initial transient. The first numerical evidence for the existence of a complexity-condensed state is provided by the probability distribution of the complexity field for a fixed value of $\beta$ and varying values of $\alpha$.

Fig~(\ref{fig:3}) illustrates this behavior for $\beta=10$, $30$, and $50$. For each value of $\beta$, the complexity distributions are shown for a low value of $\alpha$ (blue) and for the critical value $\alpha_c$ (yellow), at which the complexity variance is minimized and the system reaches the condensed state. The corresponding variance values are shown in the inset. In addition to the narrowing of the distribution, the results reveal a shift toward a characteristic complexity value $\lambda_c$, as defined in Eq.~(\ref{eq13}). The condensed distribution is centered around a comparatively large finite-time Lyapunov exponent (FTLE), indicating the stabilization of a high-complexity regime. Thus, complexity condensation represents a homogenization of chaotic dynamics at the level of information production, rather than a reduction of chaos or the onset of microscopic state synchronization.

\begin{figure*}[t]
    \centering
    \begin{subfigure}[t]{0.3\textwidth}
        \centering
        \includegraphics[width=\linewidth]{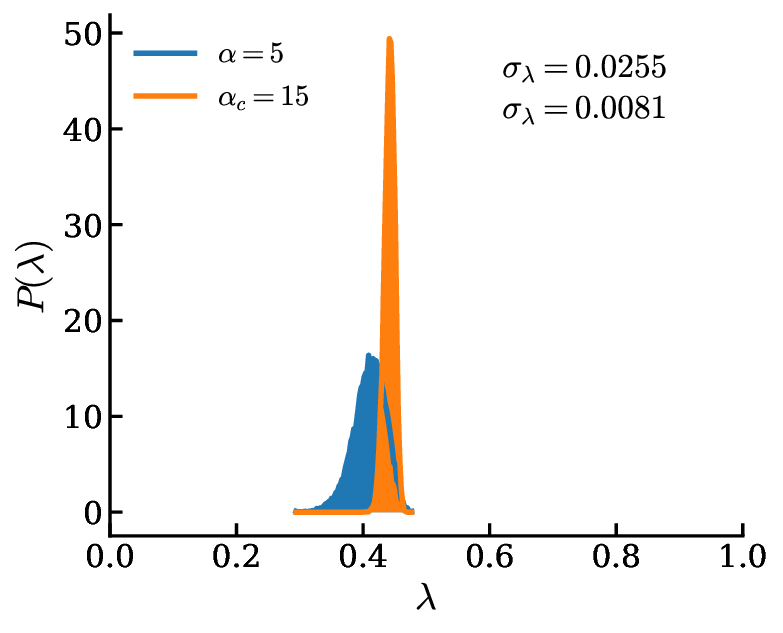}
        \caption{$\beta=10$}
        \label{fig:sub1}
    \end{subfigure}
    \hfill
    \begin{subfigure}[t]{0.3\textwidth}
        \centering
        \includegraphics[width=\linewidth]{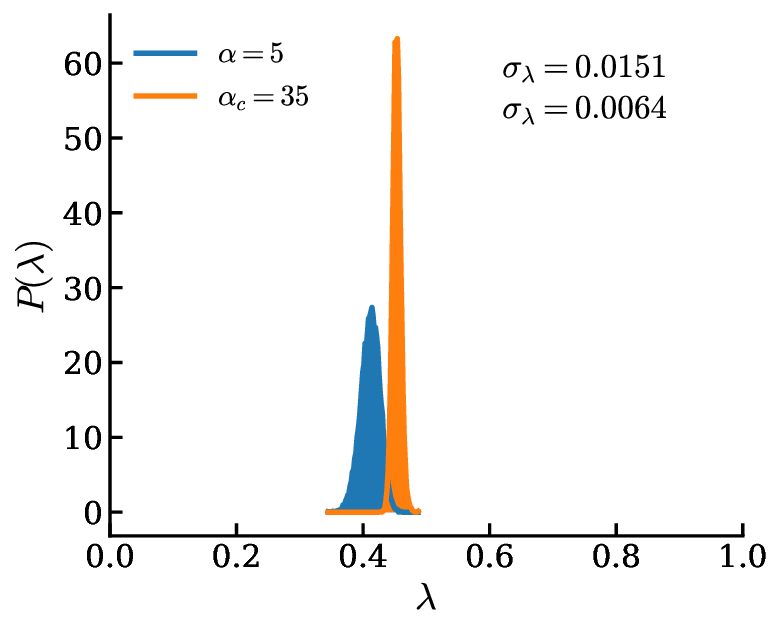}
        \caption{$\beta=30$}
        \label{fig:sub2}
    \end{subfigure}
    \hfill
    \begin{subfigure}[t]{0.3\textwidth}
        \centering
        \includegraphics[width=\linewidth]{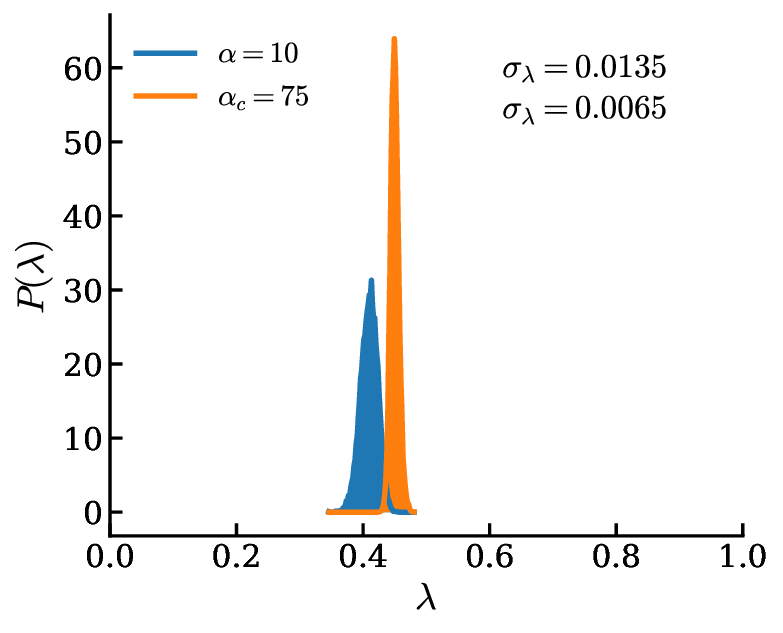}
        \caption{$\beta=50$}
        \label{fig:sub3}
    \end{subfigure}

    \caption{$P(\lambda)$ Vs. $\lambda$ for three different adaptive sensitivities $\beta$. Increasing adaptive selectivity decreases the variance and narrows the complexity distribution, indicating complexity condensation.}
    \label{fig:3}
\end{figure*}
\begin{figure*}[t]
    \centering
    \begin{subfigure}[t]{0.3\textwidth}
        \centering
        \includegraphics[width=\linewidth]{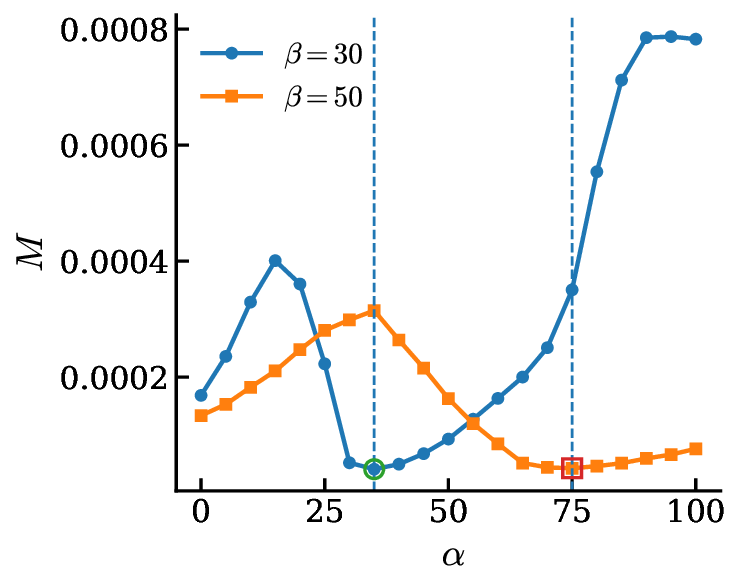}
        \caption{$M$  vs. $\alpha $}
        \label{fig:sub4}
    \end{subfigure}
    \hfill
    \begin{subfigure}[t]{0.3\textwidth}
        \centering
        \includegraphics[width=\linewidth]{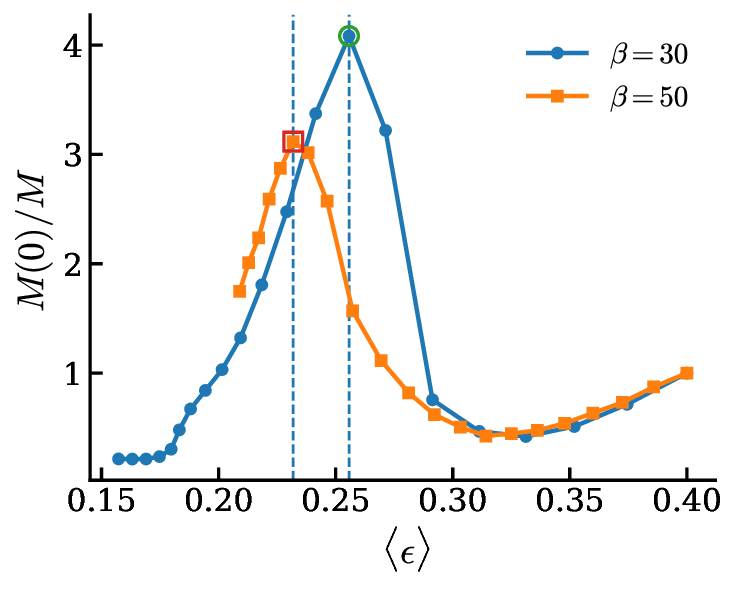}
        \caption{$\frac{M(0)}{M}$ Vs. $\alpha$}
        \label{fig:sub5}
    \end{subfigure}
    \hfill
    \begin{subfigure}[t]{0.3\textwidth}
        \centering
        \includegraphics[width=\linewidth]{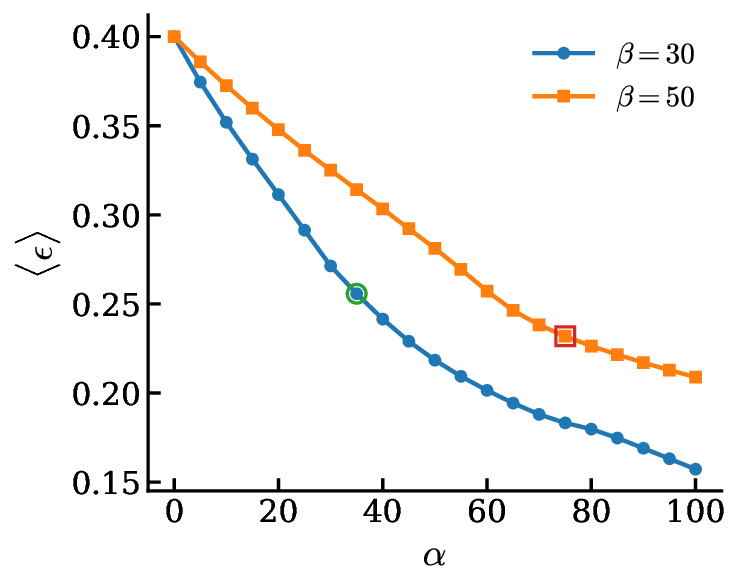}
        \caption{$<\epsilon>$ Vs. $\alpha$}
        \label{fig:sub6}
    \end{subfigure}

    \caption{Dependence of the complexity measures on the adaptive feedback strength $\alpha$. (a) Complexity variance, $M(\alpha)$, as a function of $\alpha$. (b) Complexity condensation efficiency, $\eta=M(0)/M(\alpha)$, illustrating the relative reduction in the variance of the complexity measure with respect to the reference state.
(c) Mean adaptation cost, $\langle\epsilon\rangle$, as a function of $\alpha$. The results demonstrate how increasing adaptive feedback simultaneously modifies the complexity variance, enhances complexity condensation, and influences the average adaptation cost.}
    \label{fig:4}
\end{figure*}

Numerical simulations reveal a nonmonotonic dependence of the complexity variance $M$ on the adaptive sensitivity parameter $\alpha$, as shown in Fig.~(\ref{fig:sub4}). For a fixed value of $\beta$, the variance initially increases as $\alpha$ is increased from zero, indicating an enhancement of complexity heterogeneity across the network. Beyond a threshold value of $\alpha$, however, a sharp decrease in $M$ is observed, suggesting that adaptive filtering suppresses interactions between dynamically incompatible subsystems and drives the system toward a condensed state in complexity space. Further increases in $\alpha$ gradually weaken information exchange, leading to a loss of condensation and a corresponding increase in the variance. Consequently, the maximum degree of complexity condensation occurs at an intermediate adaptation strength $\alpha_c$(35 for $\beta=30$ and 75 for $\beta=50$), where $M$ attains its minimum value. This optimum reflects a balance between information mixing and dynamical isolation.
As $\beta$ is increased, the qualitative behavior remains unchanged, but the location of the optimal adaptation strength shifts toward larger values of $\alpha$. This observation indicates that complexity condensation arises from the interplay of two complementary mechanisms. The first is the structural selectivity controlled by $\beta$, which restricts information exchange to dynamically similar subsystems in complexity space. The second is the adaptive filtering controlled by $\alpha$, which suppresses the incorporation of information from complexity-mismatched neighborhoods. Increasing $\beta$ enhances structural selectivity and therefore requires a stronger adaptive sensitivity to achieve optimal condensation. As a result, the condensation minimum shifts to higher values of $\alpha_c$ with increasing $\beta$.

From the perspective of the emergent coupling strength, maximal complexity condensation occurs at an intermediate level of information exchange as shown in Fig~(\ref{fig:sub5}). The degree of condensation is reduced for both excessively strong and excessively weak coupling, whereas an optimal coupling window produces the strongest localization of the complexity field. This behavior indicates that complexity condensation is not governed by maximal communication alone, but rather by a balance between information exchange and adaptive selectivity. Furthermore, the optimally condensed state is realized at progressively smaller effective coupling strengths, suggesting that maximal organization can emerge from relatively weak information exchange when accompanied by strong selective filtering. Complexity condensation therefore arises from an optimal level of selective information exchange and emerges neither in the limit of maximal communication nor in the limit of complete isolation.

The dependence of the mean adaptive coupling strength on the sensitivity parameter $\alpha$ for two interaction locality values, $\beta=30$ and $\beta=50$, is shown in Fig.~(\ref{fig:sub6}). In both cases, the mean coupling strength decreases approximately exponentially with increasing $\alpha$, indicating that adaptive feedback progressively suppresses interactions between dynamically incompatible units. However, the curve corresponding to the larger locality parameter ($\beta=50$) lies consistently above that for $\beta=30$. This behavior suggests that increasing the interaction locality enhances the localization of the complexity landscape, thereby reducing the average complexity mismatch between a node and its neighborhood. As a result, the adaptive penalty is weakened, allowing larger effective coupling strengths to be sustained. These results demonstrate that interaction selectivity not only suppresses communication between dynamically incompatible units but also promotes stronger interactions among dynamically compatible units, leading to a more efficient self-organized network structure.

\begin{figure}[t]
\centering
\includegraphics[width=\columnwidth]{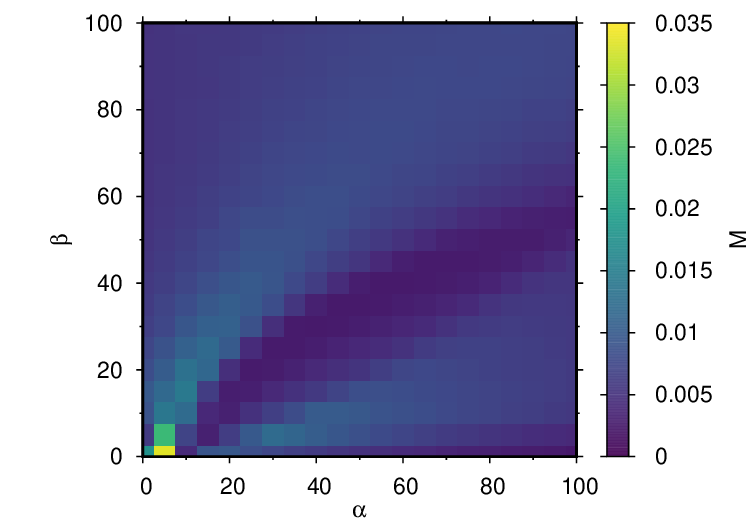}
\caption{Phase diagram of the complexity variance $M=\mathrm{Var}(\lambda)$ in the $(\alpha,\beta)$ parameter plane. Darker colors correspond to smaller values of $M$,
indicating stronger complexity condensation,whereas brighter colors represent distributed complexity states. Number of maps 100 and $T_\lambda=100$.
}
\label{fig:3}
\end{figure}

The phase diagram in the \((\alpha,\beta)\) plane shows that complexity condensation occupies a well-defined region of parameter space, rather than occurring across the entire range. This supports the view that complexity condensation is an emergent phenomenon requiring a balance between interaction selectivity and adaptive feedback strength. Fig~\ref{fig:3} displays the condensation order parameter \(M\) in the \((\alpha,\beta)\) plane. A pronounced valley of minimum variance separates two regimes: condensed and distributed complexity. This valley indicates an optimal balance between adaptive coupling and complexity-selected interactions that is necessary for collective organization. As \(\beta\) increases, the location of the minimum shifts toward larger \(\alpha\), implying that stronger adaptive regulation is required to compensate for increasingly localized interaction neighborhoods. The resulting locus of optimal condensation therefore demonstrates that complexity condensation is governed by adaptive dynamics and interaction topology.

\section{Discussion}

The results demonstrate a novel form of collective organization generated by adaptive information exchange, namely the condensation of an emergent complexity field. In this phenomenon, dynamically distinct subsystems converge toward a common level of information production without undergoing complete state synchronization. Unlike conventional synchronization, where microscopic trajectories become identical, complexity condensation occurs at the level of dynamical complexity. The resulting state is characterized by a pronounced reduction in the complexity variance and a strong localization of the complexity distribution, indicating the emergence of collective organization in complexity space.

The existence of an optimal adaptation regime is one of the central findings of this study. The complexity variance exhibits a nonmonotonic dependence on the adaptive sensitivity parameter $\alpha$, producing a well-defined condensation optimum for a fixed value of the locality parameter $\beta$. For weak adaptation, dynamically dissimilar subsystems undergo excessive mixing, leading to increased complexity heterogeneity. In contrast, strong adaptation suppresses information exchange and promotes dynamical isolation. As a result, maximal complexity condensation emerges from a balance between adaptive selectivity and information sharing. From the perspective of the emergent coupling strength, the optimal condensed state is realized at an intermediate level of information exchange rather than under conditions of maximal communication. This observation indicates that complexity condensation is governed not by the quantity of information exchanged, but by the efficiency with which information is selectively filtered and incorporated into the dynamics.

The present work may have implications for a broad class of adaptive complex systems. In many real-world settings, including neural, social, and distributed computational networks, effective collective organization requires selective rather than indiscriminate information exchange. The results suggest that adaptive regulation based on local complexity differences can drive large populations toward coherent complexity states while preserving microscopic diversity. From this perspective, dynamical complexity functions not merely as a descriptor of system behavior but as an active organizing field that shapes interaction patterns and collective dynamics. The proposed framework therefore provides a new perspective on self-organization in adaptive nonlinear systems and highlights the role of complexity-dependent interactions in the emergence of large-scale organization.
\section{Conclusion}

The present work introduces an adaptive dynamical-network framework in which an emergent complexity field governs both the interaction topology and the exchange of information. The network self-organizes in complexity space through a feedback loop linking microscopic dynamics, complexity-dependent interactions, and adaptive coupling. Dynamical complexity is quantified using finite-time information production rates, which serve as the basis for both interaction selection and adaptive regulation.
Numerical simulations reveal the emergence of complexity condensation, characterized by a strong localization of the complexity distribution even in the absence of microscopic synchronization. We identify an optimal regime of information exchange in which condensation is maximized at an intermediate coupling strength. The location of this optimum shifts systematically with the complexity-space locality parameter, reflecting the interplay between structural selectivity and adaptive regulation. These findings demonstrate that emergent complexity fields can act as active organizing agents in adaptive nonlinear networks and suggest a general mechanism through which adaptive information exchange can generate collective organization in complex systems.

\bibliographystyle{apsrev4-2}
\bibliography{reference}
\end{document}